\documentclass[aps,prb,twocolumn,superscriptaddress,showkeys,amsmath,amssymb]{revtex4-1}
\usepackage{amsmath,amssymb,upgreek}
\usepackage{graphicx}
\usepackage{bm,bbm}
\usepackage{color}
\usepackage{todonotes}
\usepackage{siunitx}
\usepackage{hyperref}

\newcommand{\GM}{$\overline{\Gamma \text{M}}$}
\newcommand{\GK}{$\overline{\Gamma \text{K}}$}
\newcommand{\Sa}{Bi$_2$Se$_3$}

\makeatletter
\newcommand\footnoteref[1]{\protected@xdef\@thefnmark{\ref{#1}}\@footnotemark}
\makeatother

\begin{document} 

\author{Adrian Ruckhofer}
\email{ruckhofer@tugraz.at}
\affiliation{Institute of Experimental Physics, Graz University of Technology, Graz, Austria}
\author{Davide Campi}
\affiliation{THEOS and MARVEL, \'{E}cole Polytechnique F\'{e}d\'{e}rale de Lausanne, Lausanne, Switzerland}
\author{Martin Bremholm}
\affiliation{Center for Materials Crystallography, Department of Chemistry and iNANO, Aarhus University, 8000 Aarhus, Denmark}
\author{Philip Hofmann}
\affiliation{Department of Physics and Astronomy, Aarhus University, Aarhus, Denmark}
\author{Giorgio Benedek}
\affiliation{Dipartimento di Scienza dei Materiali, Universit\`{a} degli Studi di Milano-Bicocca, Milano, Italy}
\affiliation{Donostia International Physics Center (DIPC) and University of the Basque Country, 20018 Donostia / San Sebastian, Spain}
\author{Marco Bernasconi}
\affiliation{Dipartimento di Scienza dei Materiali, Universit\`{a} degli Studi di Milano-Bicocca, Milano, Italy}
\author{Wolfgang E. Ernst}
\affiliation{Institute of Experimental Physics, Graz University of Technology, Graz, Austria}
\author{Anton Tamt\"{o}gl}
\email{tamtoegl@gmail.com}
\affiliation{Institute of Experimental Physics, Graz University of Technology, Graz, Austria}

\title[THz Surface Modes in Bi$_2$Se$_3$(111)]{Terahertz Surface Modes and Electron-Phonon Coupling in Bi$_2$Se$_3$(111)}

\begin{abstract}
We present a combined experimental and theoretical study of the surface vibrational modes of the topological insulator (TI) \Sa\ with particular emphasis on the low-energy region below 10 meV that has been difficult to resolve experimentally. By applying inelastic helium atom scattering (HAS), the entire phonon dispersion was determined and compared with density functional perturbation theory (DFPT) calculations. The intensity of the phonon modes is dominated by a strong Rayleigh mode, in contrast to previous experimental works. Moreover, also at variance with recent reports, no Kohn-anomaly is observed. These observations are in excellent agreement with DFPT calculations. Besides these results, the experimental data reveal---via bound-state resonance enhancement---two additional dispersion curves in the gap below the Rayleigh mode. They are possibly associated with an excitation of a surface electron density superstructure that we observe in HAS diffraction patterns. The electron-phonon coupling paramenter $\lambda$ = 0.23 derived from our temperature dependent Debye-Waller measurements compares well with values determined by angular resolved photoemission or Landau level spectroscopy. Our work opens up a new perspective for THz measurements on 2D materials as well as the investigation of subtle details (band bending, the presence of quantum well states) with respect to the electron-phonon coupling.
\end{abstract}

\maketitle 

\maketitle 	

\section{Introduction}
\begin{figure}[htb]
	\centering
	\includegraphics[width=0.48\textwidth]{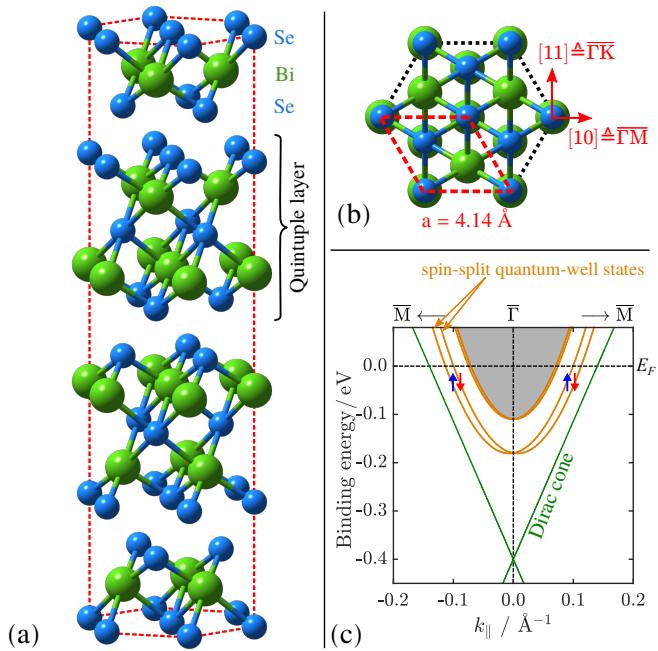}
	\caption{(a) Crystal structure of \Sa\ based on the conventional hexagonal unit cell. The unit cell consists of three quintuple layers of which each one is terminated by a Se layer. (b) Top view of the (111) surface with the red rhombus illustrating the hexagonal surface unit cell with lattice constant $a$. The two high-symmetry scanning directions are indicated by the red arrows. (c) Illustration of the electronic dispersion around the $\overline{\Gamma}$-point (see Refs. \cite{Bianchi2010,King2011}). Despite the Dirac cone, quantum-well states as illustrated by the orange lines exist on the surface as well. The quantum-well states are split due to Rashba coupling, thus making the previously degenerate spin states (illustrated by the red and blue arrows) separate into an inner and outer branch, each having a spin texture with opposite chirality.}
	\label{fig:struct}
\end{figure}

\Sa\ (\autoref{fig:struct}) is categorised as a three-dimensional topological insulator, a new state of quantum matter with a bulk gap and spin-orbit split surface states forming a Dirac cone across the gap\cite{Hasan2010,Moore2009}. The interaction of electrons with surface phonons in \Sa\ has been studied intensively due to its important role in transport properties and possible constraints for potential applications in a variety of nanotechnologies\cite{Hatch2011,Pan2012,Chen2013,Kondo2013,Zhu2012,Giraud2012,DiPietro2013}. Bismuth selenide as well as telluride alloys are classical thermoelectric materials\cite{Kadel2010,Mishra1997} with a large Seebeck coefficient and, as such, they have been used in thermoelectric refrigeration for a long time\cite{Springer1964}. However, to fully understand the thermoelectric properties of \Sa\ thin films and nanoscale devices\cite{Liang2016,Tang2015}, information on the surface phonon dispersion curves and the electron-phonon (e-ph) interaction is crucial\cite{Liang2016,Minnich2009,Hsiung2015}.\\ 
So far experimental information about the surface phonon dispersion curves of \Sa(111) was limited to previous helium atom scattering (HAS) studies by Zhu \emph{et al.}\cite{Zhu2011,Zhu2012}, in the low energy part of the phonon spectrum. These studies suggested the presence of a deep Kohn anomaly in the 7.5 meV optical phonon branch (S2) coupled to the electronic (spin-forbidden) transition across the Dirac cone\cite{Zhu2011}. However, existing first-principle calculations of \Sa(111) phonon dispersion curves, do not show any evidence of Kohn anomalies in the S2 branch\cite{Heid2017}.\\
A convenient parameter to characterise the e-ph coupling strength is the mass-enhancement $\lambda$\cite{Grimvall1981} and in recent years it was demonstrated that HAS from conducting surfaces can directly measure the mode-selected e-ph coupling constants $\lambda_{\mathbf{Q}j}$\cite{Sklyadneva2011,Tamtoegl2013}, besides the surface phonon dispersion curves\cite{BenedekTB}. Moreover the temperature-dependence of the HAS Debye-Waller factor was shown to provide the global e-ph coupling constant $\lambda$ at the surface of thin metal films\cite{Benedek2018,BenedekTB} and topological insulators\cite{Tamtogl2017b}. Yet the large $\lambda$ as estimated based on the Kohn anomaly\cite{Zhu2012} is at odds with theoretical findings that indicate that the major contribution to $\lambda$ comes from the higher optical phonon branches\cite{Heid2017}, especially when the Fermi level crosses the surface quantum-well states above the conduction band minimum (see \autoref{fig:struct}(c)). The situation is actually met in recent high-resolution $^3$He-spin scattering studies on Bi$_2$Te$_3$(111), where the weak signature of a Kohn anomaly is detected in the surface longitudinal acoustic resonance\cite{Tamtogl2018a}, also not found in adiabatic \emph{ab-initio} calculations of the phonon branches.\\
In order to elucidate these conflicting results, we have undertaken a HAS study of the surface phonon dispersion curves and the e-ph interaction of \Sa(111). Supersonic neutral He atom beams with incident energies in the range $\leq 20~\mbox{meV}$ have been used to probe low-energy surface excitations with the best available resolution, while being an inert completely nondestructive probe\cite{Farias1998,BenedekTB}. The technique allows to measure most of the surface phonon branches in the acoustic and optical regions. Low-energy He atoms impinging on a conducting surface are exclusively scattered by the surface charge density\cite{Mayrhofer2013,Tamtoegl2013} and inelastic scattering from surface phonons only occurs via the phonon-induced charge density oscillations, i.e., via the e-ph interaction. It is in this way that inelastic HAS provides a first-hand information on the e-ph interaction, with the neutral He atoms acting as a sort of local mechanical probe on the electron density.\\
Energy and momentum, inelastically exchanged by He atoms with the surface can, however, be retained by the electron system in the form of low-energy collective excitations. In principle, the HAS signal from this kind of excitations is expected to be quite small. Nevertheless, an increased e-ph interaction due to surface quantum-well states\cite{Silkin2004} in combination with an enhancement from HAS bound-state resonances\cite{Evans1983}, suggests to assign two branches of low-energy modes in the gap well below the Rayleigh waves (RW) to some sort of collective electronic excitations. Actually anomalous acoustic plasmons have been recently reported in \Sa(111) by Jia \emph{et al.}\cite{Jia2017}, from high-resolution electron energy-loss spectroscopy, although these modes turn out to be superimposed in the first Brillouin zone onto the RW branch.\\
Plasmons in a two-dimensional electron gas (2DEG) with a $\sqrt{\mbox{Q}}$ dispersion (2D plasmons) have been predicted long ago by Frank Stern\cite{Stern1967,Ando1982}. Later it was shown that the coupling of 2DEG plasmons arising from two different quantum-well minibands, as found in semiconductor surface accumulation layers, yield a surface plasmon pair: a 2D plasmon and an acoustic surface plasmon (ASP) with a linear dispersion above the upper edge of the single-particle excitation spectrum\cite{Shvonski2019,Hermanson1990}. Similarly the coupling of a 2DEG at a metal surface coupled to the underlying 3D  electron gas yields an ASP in addition to the ordinary surface plasmon\cite{Silkin2004,Pitarke2006,Wang2011,Diaconescu2007}. As discussed below, the assignment of the two additional low-energy branches as collective polaron excitation recently suggested by Shvonski \emph{et al.}\cite{Shvonski2019}, although plausible in semimetals with a large dielectric constant, definitely requires further ad-hoc studies, possibly with even higher resolution as available, e.g., with $^3$He spin-echo spectroscopy\cite{Tamtogl2018a}.
	
\section{Results and discussion}
\begin{figure}[htb]
	\centering
	\includegraphics[width=0.49\textwidth]{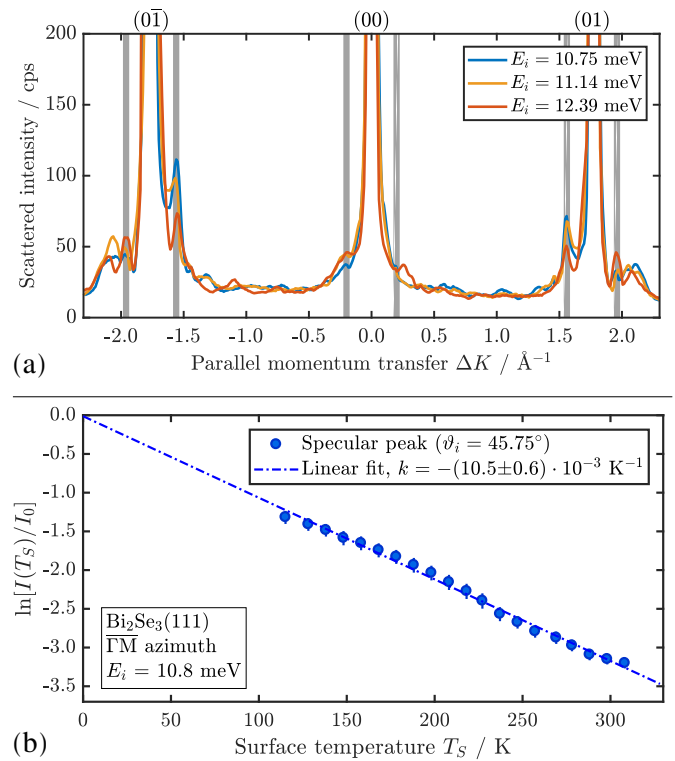}
	\caption{(a) HAS diffraction scans for \Sa(111) measured at various incident energies with the sample at 113 K, aligned along the \GM\ azimuth. The smaller features between the specular and the two diffraction peaks are due to bound-state resonances and kinematic focusing\cite{Ruckhofer2019}. The vertical shaded regions next to the diffraction peaks illustrate additional features which appear to be independent of $E_i$, at a distance of about $0.2~\mbox{\AA}^{-1}$ form the specular and first order diffraction peaks. (b) The temperature dependence of the Debye-Waller exponent of \Sa(111) for the specular peak with the sample aligned along \GM\ .}
	\label{fig:diffraction}
\end{figure}
\autoref{fig:diffraction}(a) shows several diffraction scans along the \GM\ azimuth plotted versus momentum transfer $\Delta K$, while in \autoref{fig:diffraction}(b) the temperature dependence of the specular peak intensity is plotted, which can be used to extract the electron-phonon coupling constant $\lambda$ (section \nameref{sec:EPhCoupling}). The diffraction scans in \autoref{fig:diffraction}(a) have been measured at three different incident energies $E_i$, at a sample temperature of 113 K. The intensity scale has been scaled to show additional features with smaller intensity. Besides some features assigned to bound-state resonances and kinematic focusing, which are easily recognised due to the strong dependence of their position on the incident energy\cite{Mayrhofer2013}, there are features that occur at fixed values of $\Delta K$, independently of $E_i$ with a distance of about $0.2~\mbox{\AA}^{-1}$ from the specular and first order diffraction peaks, as indicated by the vertical shaded regions (further diffraction scans, including also the \GK\ azimuth can be found in the supplementary information).\\
We recently observed with HAS a multivalley charge density wave (CDW) in Sb(111) originating from the $\overline{\mathrm{M}}$-point electron pockets giving rise to additional peaks in the diffraction pattern\cite{Tamtogl2019}. In \Sa(111), however, no carrier pockets exist besides the Dirac cone and the quantum-well minibands occurring around the zone center (see \autoref{fig:struct}(c)). The latter provide nesting wavevectors of about $0.2~\mbox{\AA}^{-1}$ between states of equal spin, which correspond fairly well to the parallel momentum transfers of the satellites observed aside the $(0,\pm 1)$ peaks in HAS diffraction spectra (and likely also aside the specular peak, despite the coincidence with bound-state resonances) (\autoref{fig:diffraction}(a)). It should be noted, however, that the observation of satellite peaks whose position is independent of the HAS incident energy is by itself indicative of a long-period superstructure of the electron density, possibly incommensurate or weakly commensurate with the surface atomic lattice. Charge density oscillations as low as $10^{-6}$ atomic units, presently accessible to HAS, can in principle sustain very low-energy collective phase and amplitude excitations in the meV spectral range\cite{Tamtogl2019}, and possibly suggest an assignment of the present low-energy branches.\\
	
\subsection{Time-of-flight measurements and phonon dispersion curves}
\label{sec:phonondisp}
\begin{figure}[!ht]
	\centering
	\includegraphics[width=0.45\textwidth]{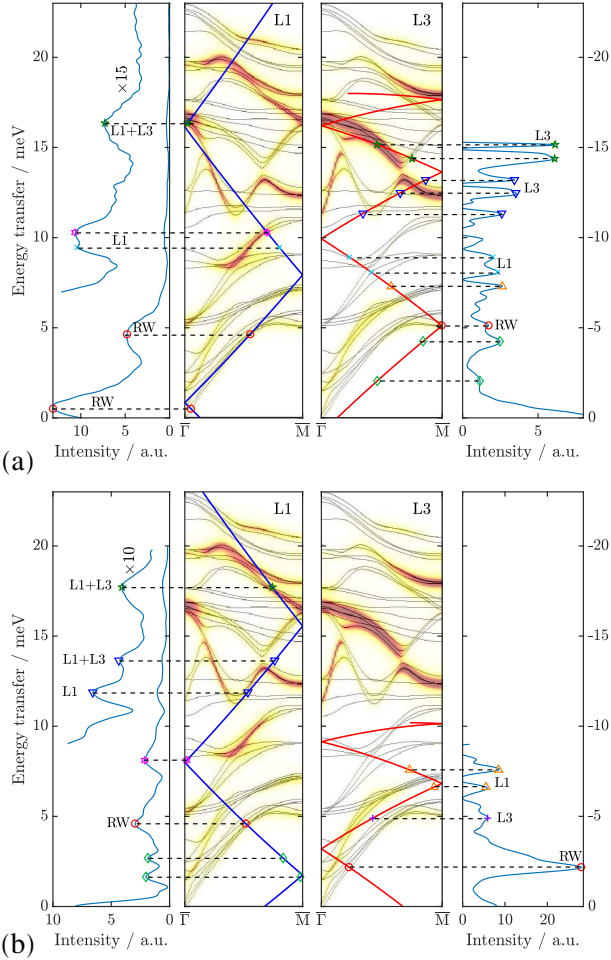}
	\caption{Plot of two typical TOF measurements with the sample along \GM\ and at room temperature. In (a) the incident energy is $E_i = 17.97~\mbox{meV}$ with $\vartheta_i = 34.25^{\circ}$ while in (b) $E_i = 10.19~\mbox{meV}$ and $\vartheta_i = 40.25^{\circ}$. The left and right-most panel show the TOF spectrum after conversion to energy transfer. The blue (phonon annihilation, $\Delta E > 0$) and red (phonon creation, $\Delta E < 0$) lines in the centre panels show the scan curves superimposed onto the calculated dispersion (The colour code giving the intensity of the longitudinal L1 and L3 modes projected onto the first and third surface layer, respectively, is that of \autoref{fig:DFPTSimulationen}). The symbols denote peaks in the TOF spectrum which have been assigned to phonon events (RW = Rayleigh wave). The two distinct low-energy peaks (green diamonds) observed on the creation (panel (a)) and annihilation (panel (b)) sides in the phonon gap below the RW are tentatively assigned to low-energy collective surface electronic excitations, associated with the long-period surface superstructure and revealed through HAS bound-state resonance enhancement.}
	\label{fig:TOF}
\end{figure}

\begin{figure}[htb]
	\centering
	\includegraphics[width=0.5\textwidth]{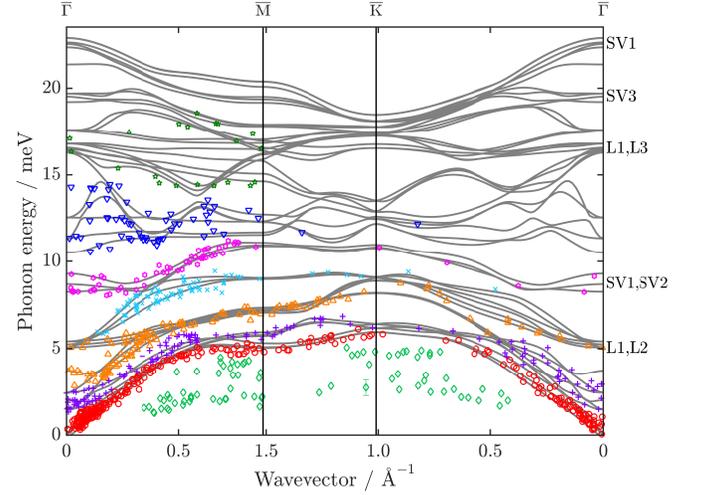}
	\caption{Comparison of the measured phonon dispersion relation with DFPT calculations for three quintuple layers. The solid grey lines represent the DFPT calculation while different symbols for the experimentally determined points correspond to different phonon modes. The assignment of the symbols is based on the proximity to particular modes.}
	\label{fig:PhononDisp}
\end{figure}

\begin{figure*}[!ht]
	\centering
	\includegraphics[width=0.7\textwidth]{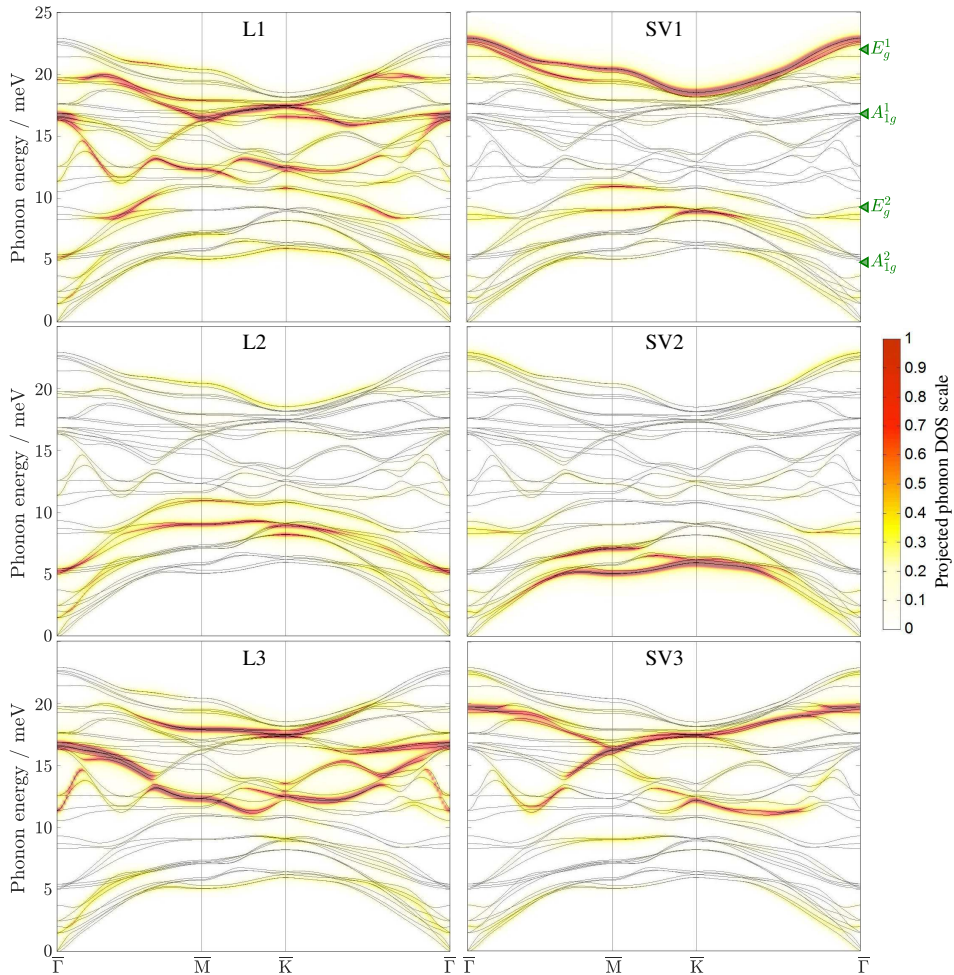}
	\caption{Polarisation of the calculated phonon dispersion of \Sa\ from DFPT. The left column shows longitudinal polarisation for the topmost three atomic layers (L1-L3). The right column shows shear vertical polarisation (SV1-SV3). The colour code bar on the right-hand side gives the intensity of the mode projected onto the corresponding layer. The green triangles at the top-right axis indicate the experimental energies of bulk Raman active modes according to Ref. \cite{Boulares2018}.}
	\label{fig:DFPTSimulationen}
\end{figure*}

The phonon energies were determined by performing time-of-flight (TOF) measurements over a wide range of incident angles between the first-order diffraction peaks and at various beam energies. The phonon dispersion was then obtained by calculating the parallel momentum transfer $|\Delta \mathbf{K}|$ for each extracted phonon energy from the conservation laws of energy and parallel momentum providing the so-called scan curve for planar scattering (see \eqref{eq:scancurve} and Refs. \cite{Tamtoegl2013,Safron1996}).\\
In \autoref{fig:TOF}(a) an example of a TOF spectrum after conversion to the energy transfer scale is shown. The measurement was taken in the high symmetry direction $\overline{\Gamma \text{M}}$ with an incident beam energy $E_i = 17.97~\mbox{meV}$ and at an incident angle of $\vartheta_i = $ \SI{34.25}{\degree}. The TOF spectrum consists of several peaks which are located on the creation ($\Delta E<0$) as well as the annihilation ($\Delta E>0$) side. The peak at zero energy transfer corresponds to elastically scattered helium atoms\cite{Farias1998,Tamtogl2018a}. The scan curve, shown in the two centre panels of \autoref{fig:TOF} for phonon annihilation (blue) and creation (red) events, associates each phonon event with a specific momentum transfer $\Delta K$. The scan curve has been backfolded into the irreducible part of the Brillouin zone and is plotted on top of the calculated dispersion. The different symbols on the scan curves, marking the main inelastic features, have been associated to phonons of different character and polarisation.\\
The large peaks in the TOF spectra marked with the red circles in \autoref{fig:TOF}, correspond to the Rayleigh wave (RW) as seen in the DFPT calculations. Note that in the present TOF spectra the RW exhibits typically the largest intensity of all inelastic events (cfr. the intensities in \autoref{fig:TOF}(a)). There is a fair correspondence between the present HAS data and those previously reported by Zhu \emph{et al.}\cite{Zhu2011}. Curiously Zhu \emph{et al.} stated that the RW is not observed, whereas it appears in their plot, though with only a few data points, in reasonable agreement with the present one in the \GK\ direction; it also occurs in the \GM\ direction, once it is recognised that there is an avoided crossing, so that the RW at $\overline{\text{M}}$ is not the lowest mode. There is however an important difference with respect to Zhu \emph{et al.}\cite{Zhu2011}: Present HAS data do not show any evidence of a Kohn anomaly in the $\approx 8$ meV branch for wavevectors around $0.2~\mbox{\AA}^{-1}$ and associated with the nesting at the Fermi level across the Dirac cone (or more likely across the parabolic dispersion of surface quantum-well states\cite{King2011,Bianchi2010}).\\
\autoref{fig:PhononDisp} shows the entire experimental surface phonon dispersion (symbols) superimposed onto the DFPT calculations (grey lines). The different symbols have been associated to phonons of different character and polarisation based on the proximity to particular modes of the DFPT calculations. In total, we are able to distinguish at least 8 different branches.\\
The polarisation analysis of the calculated surface phonon modes can be found in \autoref{fig:DFPTSimulationen} where the intensity of each polarisation projected onto the corresponding layer is given by the colour code. The left column shows the longitudinal polarisations for the first (L1), second (L2), and third (L3) layer. The right column shows the shear vertical polarisation for the first three layers (SV1-SV3), while the shear horizontal polarisation can be found in the supporting information. The theoretical dispersion curves are seen to agree quite well with the HAS data and also with the experimental Raman-active modes at $\overline{\Gamma}$ (green triangles in \autoref{fig:DFPTSimulationen} according to\cite{Boulares2018}).\\ 
A closer comparison of the experimental data points in \autoref{fig:PhononDisp} with \autoref{fig:DFPTSimulationen} shows that mainly phonon events with the largest amplitude in the two topmost layers of the sample are observed in the experiment. In particular in the low energy region ($< 10$ meV), most contributions come from phonons with the largest amplitude in the second layer (L2, SV2), which is a Bi layer and therefore about $2.5$ times heavier than the first Se layer. The most prominent mode in the TOF spectra, the RW, corresponds predominantly to L and SV polarisations, due to its elliptical polarisation, with a particularly strong SV2 component at the Brillouin zone boundary.\\
Note that the calculation does not reproduce the Kohn anomaly reported by Zhu \emph{et al.}\cite{Zhu2011} (see the SI for calculations with SOC), in agreement with other recent ab-initio calculations including SOC by Heid \emph{et al.}\cite{Heid2017} as well as with the experimental and theoretical studies of Bi$_2$Te$_3$(111)\cite{Tamtogl2018a}. Heid \emph{et al.}\cite{Heid2017} suggested that the Kohn anomaly observed by Zhu \emph{et al.}\cite{Zhu2011,Zhu2012} may be connected to a strong e-ph interaction in the doped bulk material rather than to a surface state. They actually showed that the largest contribution to the e-ph coupling comes from an in-plane polar-type branch in the optical region between 10 and 18 meV\cite{Heid2017}.
Indeed the anomalously strong dispersion of the optical branches in that region (\autoref{fig:DFPTSimulationen}), also found in Bi and Sb tellurides\cite{Tamtogl2018a,Campi2018} may be regarded as a manifestation of e-ph interaction.\\
In the acoustic region and the long wavelength limit (close to $\overline{\Gamma}$) the dispersion relation of the RW is linear. Its slope provides the RW group velocity in the two symmetry directions \GM\ and \GK :
\begin{align*}
	\overline{\mathrm{\Gamma M}}: \; & v_{RW} (112) = (1.63 \pm 0.07)~\mbox{km}/\mbox{s}  \\
	\overline{\mathrm{\Gamma K}}: \; & v_{RW} (110) = (1.80 \pm 0.15)~\mbox{km}/\mbox{s} \; .
\end{align*}
In order to appreciate the degree of localisation of the RW in the two symmetry directions, these values are to be compared with the corresponding speeds of sound. The present DFPT values (compared with values in parentheses derived from the available elastic constants\cite{Gao2016}) are:
\begin{align*}
	\overline{\mathrm{\Gamma M}}: \quad & v_{T,SV}  = 1.757~(1.91)~\mbox{km}/\mbox{s} \quad  v_{T,SH}  = 2.290~ (2.24)~\mbox{km}/\mbox{s} \\
	\overline{\mathrm{\Gamma K}}: \quad & v_{T,SV}  = 3.227~(2.93)~\mbox{km}/\mbox{s} \quad  v_{T,SH}  = 1.416~ (1.22)~\mbox{km}/\mbox{s} \; .
\end{align*}
It appears that the RW has a velocity in the \GM\ direction smaller than both transverse bulk values and is therefore a localised surface wave, whereas in the \GK\ direction it has a velocity larger than that of the SH transverse sound, and is therefore a pseudo-surface wave (PSW)\cite{Farnell1970,Farnell1978}. Actually in the absence of mirror symmetry for the sagittal plane in this direction, the RW is a resonance. The fact may have suggested (see Zhu \emph{et al.}\cite{Zhu2011,Zhu2012}) that in \Sa(111) the RW is suppressed but the present comparison with the DFPT calculation confirms that the RW is actually observed in both directions, though as a resonance along \GK. Values for the bulk longitudinal ($v_{L}=2.9~\mbox{km}/\mbox{s}$) and transverse ($v_{T}=1.7~\mbox{km}/\mbox{s}$) group velocities of \Sa\ have also been reported in the framework of the isotropic elastic continuum theory\cite{Giraud2012,Glinka2015b}. In this approximation the corresponding RW velocity, obtained by solving the cubic Rayleigh equation,\cite{Kress1991} would be $v_{RW}=1.56~\mbox{km}/\mbox{s}$ in any direction. 
	
\subsection{Low-energy branches}
\label{sec:plasmons}
The measured HAS-TOF spectra displayed in \autoref{fig:TOF} show also distinct peaks yielding two branches of elementary excitations with an energy below the RW branch (green diamonds in \autoref{fig:PhononDisp}). On the basis of present DFPT surface phonon dispersion curves, they cannot be attributed to any possible phonon branch of the ideal surface. HAS from conducting surfaces exclusively occurs through the interaction (mostly Pauli repulsion) with the surface electron density, and therefore also electronic excitations in the THz range can be observed by HAS, with a $0.5~\mbox{meV}$ resolution and sensitivity to charge density oscillations in the $10^{-6}$ atomic units range. \\
Actually the observed low-energy branches are reminiscent of those recently observed with HAS in Sb(111), which have been attributed to elementary excitations (phasons/amplitons) of a multi-valley CDW\cite{Tamtogl2019}. The concomitant presence of a commensurate component associated with the $\overline{\mathrm{M}}$-point electron pockets at the Fermi level, and an incommensurate one due to the hole pockets along the \GM\ direction, allows for collective excitations with a comparatively small gap at $Q = 0$. On the other hand no low-energy phason/ampliton modes have been detected with HAS for the perfectly commensurate multivalley CDW reported in the quasi-1D surface Bi(114)\cite{Hofmann2019}, discommensuration being a requisite for depinning and a vanishing/small gap at $\overline{\Gamma}$. \Sa(111) has no pocket states at the Fermi level, besides the rings around $\overline{\Gamma}$ of the surface topological Dirac and the quantum-well states\cite{King2011,Bianchi2010}. The satellites near the HAS diffraction peaks (see \autoref{fig:diffraction}(a)) suggest some long-period charge-density structures and possibly low-energy collective excitations. In order to detect the associated, seemingly small inelastic features in the TOF spectra, we rely on the bound-state resonance enhancement method (Ref. \cite{Evans1983} and Chap. 10 of Ref. \cite{BenedekTB}), applicable to highly corrugated surfaces and successfully used to detect with HAS high-energy optical surface modes in ionic crystals\cite{Bracco1986,BenedekTB}. The complete set of He-surface bound states has been measured previously\cite{Ruckhofer2019}.\\
\begin{figure*}[!ht]
	\centering
	\includegraphics[width=0.9\textwidth]{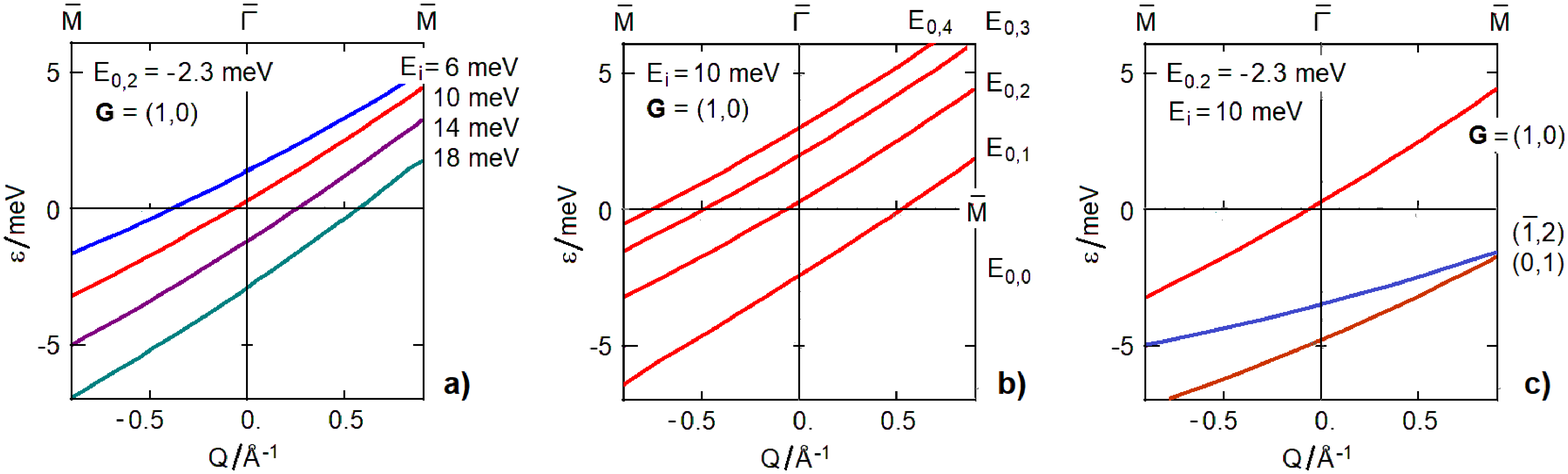}
	\caption{Inelastic bound-state resonance conditions along the \GM\ direction in the first Brillouin zone: (a) for $\textbf{G} = (1,0) \equiv (1.76,0)~\mbox{\AA}^{-1}$, the bound state $n = 2$ and different incident energies; (b) for a given $E_i = 10$ meV, \textbf{G} = $(1,0)$ and the different bound states $n = 0-4$; (c) for $E_i = 10$ meV, the bound state $n = 2$ and three different \textbf{G} vectors $(1,0), (\bar{1},2) \equiv (0, 3.05)~\mbox{\AA}^{-1}$, $(0,1) \equiv (0.88,1,53)~\mbox{\AA}^{-1}$, where the integer indices refer to the hexagonal reciprocal lattice base, while the components in $\mbox{\AA}^{-1}$ units are those of $\mathbf{G}=(G_{\parallel},G_{\perp})$; other \textbf{G}-values yield inelastic resonances in the low-energy range $\left| \varepsilon\right| < 5$ meV only for wavevectors $Q$ outside the first BZ. Suitable combinations of $E_i, E_{0,n}$, and of the \textbf{G}-vector permit to scan the phonon gap region below the RW branch, so as to tune some other localised excitation and to detect it via a resonance enhancement.}
	\label{fig:PAA-SD}
\end{figure*}
Bound-state inelastic resonances occur in the HAS-TOF spectrum, with a possibly large enhancement of the inelastic intensities, at the locus of intersections of the scan curve \eqref{eq:scancurve} \cite{Tamtoegl2013} with the inelastic bound-state resonance condition \eqref{eq:PAA-SD} (see Chap. 30 of Ref. \cite{BenedekTB}). For elementary excitations with an energy $\Delta E=\varepsilon$ and wavevector $\Delta K=Q$ the equations become:
\begin{equation}
	k_i\sin\vartheta_i+Q=\sqrt{2m(E_i+\varepsilon)/\hbar}\sin\vartheta_f 
	\label{eq:scancurve} 
\end{equation}
\begin{equation}
	E_i+\varepsilon=-\left| E_{0n}\right| +\frac{\hbar^2}{2m}\left[ (k_i \sin\vartheta_i+Q+G_{\parallel})^2+G_{\perp}^2 \right] .
	\label{eq:PAA-SD} 
\end{equation}
At the mentioned intersection of \eqref{eq:scancurve} and \eqref{eq:PAA-SD}, an elementary excitation with $(\varepsilon, Q)$ assists the selective adsorption of the atom of mass $m$, incident energy $E_i$, wavevector $k_i$ and angle $\vartheta_i$ into a bound state of energy $-\left| E_{0n}\right|$, via the exchange of a surface reciprocal lattice vector $\mathbf{G}=(G_{\parallel},G_{\perp})$. On returning the \textbf{G}-vector to the surface lattice, the atom is selectively desorbed from the bound state into the final angle $\vartheta_f$. In \autoref{eq:PAA-SD} the vector \textbf{G} has been conveniently expressed via its components parallel and orthogonal to the scattering plane, respectively. In \Sa(111) the measured He-surface bound-state energies\cite{Ruckhofer2019} are $E_{0n} = -5.6 , -3.8, -2.3, -1.2, -0.5$ meV  for $n$ = 0, 1, 2, 3, 4, respectively. For a fixed scattering geometry $\vartheta_i + \vartheta_f  = \vartheta_{SD}$  (here $\vartheta_{SD}  = 91.5^\circ$), Eqs. (\ref{eq:scancurve},\ref{eq:PAA-SD}) provide, via the elimination of $\vartheta_i = \vartheta_{SD} -\vartheta_f$, the locus of intersections $\varepsilon=\varepsilon_{E_i,n;\textbf{G}}(Q)$ for any incident energy $E_i$, bound state $n$ and reciprocal surface vector \textbf{G}.\\
The three panels of \autoref{fig:PAA-SD} show some plots of $\varepsilon=\varepsilon_{E_i,n;\textbf{G}}(Q)$ in the \GM\ direction for: (a) a given bound state ($n = 2$), a \textbf{G} vector (1,0) and different values of the incident energy $E_i$; (b) a given incident energy $E_i$ = 10 meV and different bound state energies; (c) a given incident energy $E_i$ = 10 meV and bound state $n = 2$ and some different \textbf{G}-vectors, whose functions $\varepsilon=\varepsilon_{E_i,n;\textbf{G}}(Q)$ cross the phonon gap below the RW in the first BZ. In practice the phonon gap can be fully scanned by the resonance curves $\varepsilon=\varepsilon_{E_i,n;\textbf{G}}(Q)$ by varying the incident energy, so as to detect, via resonance enhancement, weak elementary excitations.\\
Since the low-energy data points appear to allign along two dispersion curves, independently of the incident energy, as well as of $n$ and \textbf{G}, rather than being spread over the entire gap, they cannot be attributed to a resonance-enhanced many-phonon background. Furthermore, frustrated translational modes of adsorbates like CO would show no dispersion and would appear at higher vibrational energies\cite{Graham2003}. More likely these points indicate two branches of low-energy excitations associated with the surface charge-density superstructure observed in the diffraction spectra, as anticipated above.\\
In this respect it is worth mentioning a recent work by Shvonski \emph{et al.}\cite{Shvonski2019} where it is argued that a strong e-ph interaction affecting the surface 2DEG of a 3D topological crystal allows for collective polaron excitations (plasmon-polarons). Their dispersion is predicted to be that of an acoustic plasmon running below the single-particle excitation spectrum as an effect of the polaron-polaron attractive interaction. The theoretical analysis by Shvonsky \emph{et al.}\cite{Shvonski2019} is actually interpreting the recent observation with high-resolution electron energy loss spectroscopy (HREELS) by Jia \emph{et al.}\cite{Jia2017} of an anomalous acoustic plasmon (AAP) mode from the topologically protected states of \Sa(111), with energy between $0$ and $6.5$ meV (and its continuation in the second zone up to $\approx 10$ meV). The present HAS data do not permit to identify this AAP due to its superposition with the RW in the first BZ and in part with other phonon branches in its continuation.
	
\subsection{Electron-phonon coupling}
\label{sec:EPhCoupling}
As shown in recent papers\cite{Tamtogl2017b,Benedek2018}, the temperature dependence of the Debye-Waller (DW) exponent plotted in \autoref{fig:diffraction}(b) permits to extract for a conducting surface the mass-enhancement parameter $\lambda$ expressing the electron-phonon coupling strength. It is related to the DW exponent by the equations:
\begin{equation}
	\begin{split}
		\lambda = \frac{ \pi } { 2 n_s } \alpha \, , \quad  \alpha & \equiv   \frac{ \phi  } { A_c \: k_{iz}^2 } \frac{ \partial \ln I(T_S )  } { k_B  \: \partial T_S} \; ,
		\label{eq:DW_lambda2} 
	\end{split}
\end{equation}
where $\phi=4.9~\mbox{eV}$ is the work function\cite{Suh2014}, $A_{c} = 14.92~\mbox{\AA}^2$ the unit cell area, $I(T_S )$ the He-beam specular intensity, $T_S$ the surface temperature, $k_{iz} = 3.18~\mbox{\AA}^{-1}$ the normal component of the incident wavevector, and $n_s$ the number of conducting layers which contribute to the phonon-induced modulation of the surface charge density\footnote{The (bulk) carrier concentration as extracted from Hall measurements of the current sample is in the region $(1.75-1.8)\cdot 10^{18}~\mbox{cm}^{-3}$ , i.e. a particularly small conductivity in the bulk suggesting that the carrier concentration at the surface may even be larger, compared to the first generation samples\cite{Bianchi2010} which had a much larger bulk charge carrier concentration.}. The latter is estimated to be $n_s = 2 \lambda_{TF} / c_0 $, where $\lambda_{TF}$ is the Thomas-Fermi screening length characterising the surface band-bending region (here $\approx 6$ nm)\cite{Bianchi2010}, $c_0 = 9.6~\mbox{\AA}$ the quintuple layers (QL) thickness, and the factor 2 indicates the 2DEG multiplicity as observed with ARPES in the current \Sa\ sample\cite{Bianchi2010}.\\
With these values and the experimental DW derivative with respect to $T_S$ from \autoref{fig:diffraction}(b), we obtain $\lambda  = 0.51$. It should be noted that, unlike in the case of low-index metal surfaces, characterised by a soft-wall repulsive potential and negligible corrugation, here the large electronic corrugation\cite{Ruckhofer2019} implies a hard-wall potential. In this case one needs to correct $k_{iz}^2$ so as to account for the acceleration impressed by the attractive part of the potential on the He atom when approaching the surface turning point (Beeby correction\cite{Farias1998}). Therefor $k_{iz}^2$ is replaced by $k_{iz}^{'2} = k_{iz}^2 + 2 m D / \hbar^2 $, where $m$ is the He mass and $D = 6.54~\mbox{meV}$ the He-surface potential well depth\cite{Ruckhofer2019}. With the Beeby correction it is found $\lambda  = 0.23$.\\
The value compares quite well with values in the literature derived from other experiments, e.g., $\lambda  = 0.25$\cite{Hatch2011}, and $\lambda  = 0.17$\cite{Chen2013} from ARPES measurements and $\lambda  = 0.26$\cite{Zeljkovic2015} from Landau level spectroscopy. A theoretical study by Giraud \emph{et al.}\cite{Giraud2012} with phonons calculated in the isotropic continuum limit gives $\lambda  = 0.42$, whereas for other ARPES measurements, where only Dirac states appear to be involved, values of $\lambda$ as low as $0.076$ to $0.088$ have been found\cite{Pan2012}.\\
From the comparison it appears that the presence of a 2DEG due to quantum-well minibands (at least two in the present analysis) plays an important role in raising the e-ph coupling strength, which is quite small when exclusively due to the Dirac topological states, to values in the range of $0.2 - 0.4$. The same conclusion follows from the theoretical analysis by Heid \emph{et al.} \cite{Heid2017}, who showed that raising the Fermi level from the Dirac point to above the conduction band minimum gives a corresponding increase of $\lambda$ from values well below $0.1$ to values in the range above $0.2$, with a substantial contribution from interband coupling and in very good agreement with the present analysis. The role of n-type doping contributing to the formation of the surface quantum-well 2DEG is quite clear in the analysis of the e-ph coupling strength in Cu-doped \Sa\ , where an analysis based on the McMillan formula \cite{McMillan1968}, indicates a value for $\lambda$ as large as $0.62$\cite{Pan2012}.
	
\section{Conclusions}
In summary, we have determined the surface phonon dispersion curves of \Sa\ along both high symmetry directions, where the largest inelastic scattering intensity is provided by the Rayleigh wave. Thus our measurements show in contrast to previous studies that the Rayleigh mode exists and is a localised surface mode in one of the high-symmetry directions (\GM ), while in the other high-symmetry direction it is actually a pseudo-surface wave (\GK ). Comparison with density functional perturbation theory calculations shows excellent agreement with the experimental data. In addition to the phonon-related losses, we observe two additional dispersion curves in the gap well below the Rayleigh mode. These two low-energy branches may correspond to collective low-energy excitations of surface electrons.\\
The appearance of these collective electronic excitations in an unprecedentedly low energy region is probably associated with a small surface charge density and an appreciable electron-phonon coupling ($\lambda  = 0.23$). However, much more detailed experiments and theoretical analysis will be needed in order to fully understand these excitations; e.g., what is the influence of the carrier concentration upon doping and what is the role of both the Dirac and the quantum-well states, with the latter providing a much larger electron-phonon interaction than the former. The analysis advocates for a more systematic study by means of elastic and inelastic HAS spectroscopy of the surface structure, the low-energy collective excitations, and the electron-phonon interaction of interesting 2D materials, where the superior space and energy resolution of HAS is hardly attainable with other current surface probes.

\section{Methods}
\subsection{Experimental Details}
The reported measurements were performed on a HAS apparatus which generates a nearly monochromatic beam $(\Delta E/E\approx 2\% )$ of $^4$He that is scattered off the sample surface in a fixed \SI{91.5}{\degree} source-sample-detector geometry. The beam is produced in a supersonic expansion of He through a $10~\mu\mathrm{m}$ nozzle followed by sampling the core of the  supersonic expansion via a $310~\mu\mathrm{m}$ skimmer. For a detailed description of the apparatus please refer to\cite{Tamtogl2010}.\\
Energy dispersive measurements for inelastic scattering can be performed using TOF measurements with a pseudo-random chopper disc. After deconvolution with the pseudo random chopper sequence, the TOF signal is further transformed to an energy transfer scale which allows to determine inelastic (phonon) scattering events\cite{Tamtogl2010}. The scattering spectra were mainly taken with the crystal at room temperature, while a few spectra were taken with the sample cooled down to \SI{115}{\K}. The incident He beam energy was varied between 10 and 18 meV.\\
The crystal structure of Bi$_2$Se$_3$ is rhombohedral, formed of QL which are bound to each other through weak van der Waals forces\cite{Michiardi2014}. The hexagonal unit cell of the Bi$_2$Se$_3$ crystal, shown in \autoref{fig:struct}, consists of 3 QLs. Each QL is terminated by Se atoms, giving rise to the (111) cleavage plane that exhibits a hexagonal structure ($a = 4.14~\mbox{\AA}$ at room temperature\cite{Chen2011}, see \autoref{fig:struct}(b)). The Bi$_2$Se$_3$ crystal was attached onto a sample holder using thermally conductive epoxy. The sample holder was then inserted into the transfer chamber using a load-lock system and cleaved \textit{in-situ}\cite{Tamtogl2016a}. The sample can be heated using a button heater on the backside of the crystal or cooled down to \SI{115}{\K} via a thermal connection to a liquid nitrogen reservoir. The sample temperature was measured using a chromel-alumel thermocouple.

\subsection{Computational Details}
The surface dynamical properties of \Sa\ were studied using DFPT calculations\cite{Baroni2001} withing the Quantum-ESPRESSO package\cite{QMEspresso}. Norm-conserving pseudopotentials and the Perdew-Burke-Ernzerhof (PBE) approximation \cite{pbe} for the exchange and correlation functional were used as implemented in the Quantum-ESPRESSO package. The surface phonon dispersion was calculated using a slab consisting of 3 QLs separated from its periodic replica by 20 \AA\ of vacuum, without the inclusion of spin-orbit corrections (SOC) (see also the SI for calculations with SOC). For an accurate calculation of the surface lattice vibrations, in principle both SOC and van der Waals (vdW) corrections are necessary, both due to the presence of heavy elements in the compound and the latter to fully account for the weak bonds between the individual quintuple layers. However, as thoroughly discussed for Bi$_2$Te$_3$(111)\cite{Tamtogl2018a}, it appears that for layered crystals with heavy elements SOC alone gives a general softening of the phonon spectrum, compensated by the inclusion of vdW correction, so that satisfactory results are obtained at a minor computational cost without both SOC and vdW corrections and with a better agreement with experiment\cite{Ortigoza2014,Tamtoegl2013}. Also for \Sa\ better agreement with the experiment is achieved with no SOC and no vdW corrections (see \autoref{sec:phonondisp}), likely due to a compensation of errors between the underbinding often characterising PBE functionals and SOC contributions and the overbinding due to vdW forces\cite{Tamtogl2018a}. More precisely, the effect of SOC was found however to be weak for the low energy surface vibrational modes of typical TIs such as Bi$_2$Te$_3$ and Sb$_2$Te$_3$\cite{Campi2018,Tamtogl2018a} while on the other hand it was shown that vdW corrections become important for an exact description of the low energy optical modes of Bi$_2$Te$_3$\cite{Tamtogl2018a}.

\section*{Associated Content}
Supporting Information accompanies this paper including additional DFPT calculations, as well as additional spectra and further details about the helium atom scattering experiments.

\section*{Acknowledgment}
We are grateful to Prof. Evgueni V. Chulkov (DIPC) and Prof. Krzysztof Kempa (Boston College) for useful discussions. The authors are grateful for financial support by the FWF (Austrian Science Fund) within the project P29641-N36 and A.T acknowledges financial support within the project J3479-N20. We would like to thank the Aarhus University Research Foundation, VILLUM FOUNDATION via the Centre of Excellence for Dirac Materials (Grant No. 11744) and the SPP1666 of the DFG (Grant No. HO 5150/1-2) for financial support. M. Bremholm acknowledges financial support from the Center of Materials Crystallography (CMC) and the Danish National Research Foundation (DNRF93).

\bibliography{literature}

\clearpage
\onecolumngrid

\renewcommand{\thepage}{\arabic{apppage}}
\pagenumbering{arabic}

\setcounter{section}{0}
\renewcommand{\thesection}{S\arabic{section}}
\section*{\large{Supplementary Information:\\ Terahertz Surface Modes and Electron-Phonon Coupling in Bi$_2$Se$_3$(111)}}

\setcounter{table}{0}
\setcounter{figure}{0}
\setcounter{figure}{0}
\makeatletter
\renewcommand{\thesection}{S\arabic{section}}
\renewcommand{\theequation}{S\arabic{equation}}
\renewcommand{\thefigure}{S\arabic{figure}}

\section{DFPT calculations and the effect of SOC}
The topologically protected Dirac cone forms a small Fermi circle around the $\overline{\Gamma}$-point which cannot be properly described with a coarse mesh and a large smearing resulting in the impossibility to capture subtle effects, such as the proposed Kohn anomaly\cite{Zhu2011} with a standard calculation. In order to verify the effect of such states on surface phonons we repeated the phonon calculations at the $q$-point corresponding to the nesting vector ($2 k_F $) by including SOC. We compared the result with that obtained by omitting the SOC.\\
\begin{figure}[htb]
    \centering
   	\includegraphics[width=0.42\textwidth]{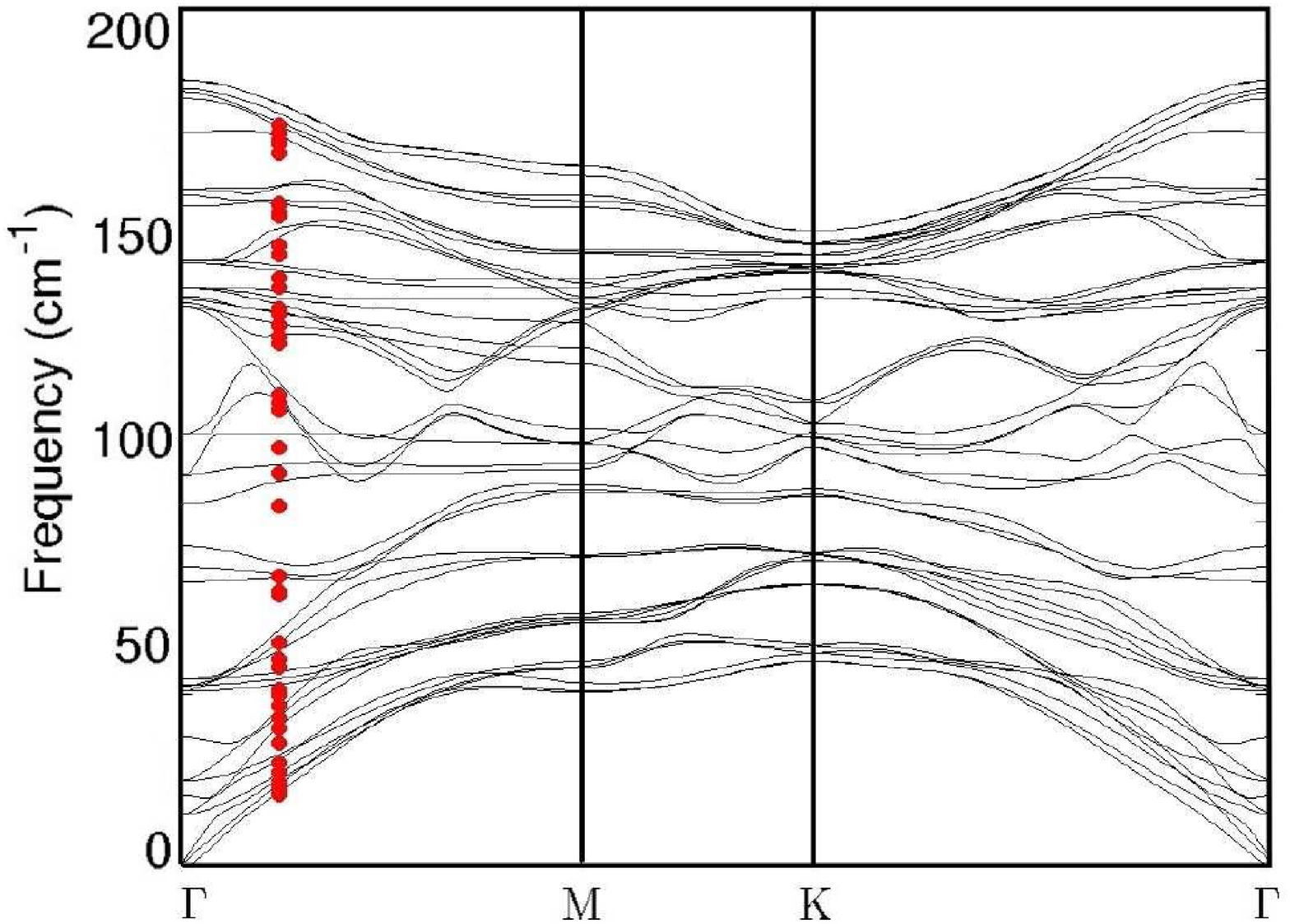}
	\caption{Surface phonon dispersion of \Sa\ omitting (black continuous line) and including spin-orbit coupling (red dots) at a $q$-vector corresponding to the nesting vector 2k$_F$. The inclusion of SOC gives rise to a softening of the phonon modes but no evidence for a Kohn anomaly is found.}
    \label{fig:DFTKA}
\end{figure}
To perform these calculations we had to improve the sampling of the Brillouin zone close to the Fermi surface which is particularly important to resolve a possibly existing anomaly. Given the
peculiar shape of the Fermi surface in the \Sa(111) slabs consisting of a ring around the $\overline{\Gamma}$-point, we used a graded $k$-point mesh (equivalent to a $50\times50\times1$ uniform mesh) near the $\overline{\Gamma}$-point point and a coarser one (equivalent to a $8\times8\times1$ mesh) near the zone boundary. The results are reported in \autoref{fig:DFTKA}. A one to one comparison between phonon modes calculated with and without SOC shows that there is no evidence of a Kohn anomaly induced by the presence of the surface metallic states, involving any of the surface phonon modes. The spin-orbit coupling results merely in a overall softening of the phonon modes of at most 6\%.\\
\begin{figure}[htb]
\centering
\includegraphics[width=0.42\textwidth]{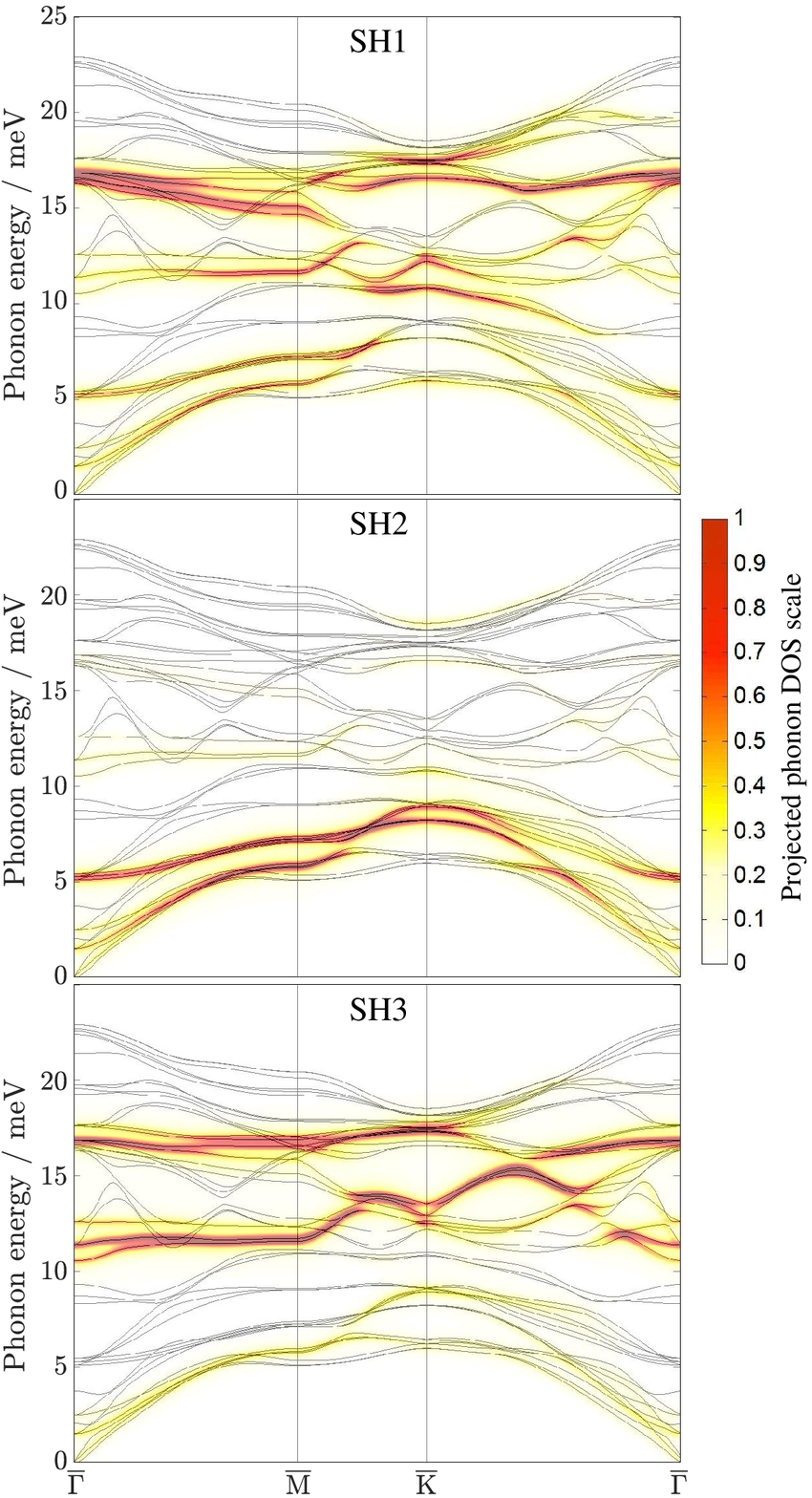}
\caption{Shear horizontal (SH) polarisation of the calculated phonon dispersion of \Sa\ for the topmost three atomic layers (SH1-SH3) from DFPT without SOC. The colour code bar on the right-hand side gives the intensity of the mode projected onto the corresponding layer.}
\label{fig:DFPTSimSI}
\end{figure}
In addition to the shear vertical and shear horizontal phonon densities shown in the main part of the article, \autoref{fig:DFPTSimSI} shows the shear horizontal polarisations projected onto the first, second and third layer (SH1, SH2, SH3). If the scattering plane, defined by the incoming and scattered He beam, coincides with a mirror plane of the surface, the detection of purely SH modes is in principle forbidden due to symmetry reasons\cite{Tamtogl2015} and we show the calculations of the SH modes here for completeness. However, phonon modes often exhibit a mixing of polarisation components and even a purely SH mode may give rise to charge density oscillations above the first atomic layer which are eventually observed in inelastic He atom scattering.\\

\section{HAS diffraction scans}
\begin{figure}[htb]
    \centering
    \includegraphics[trim=0 0 0 6.27cm, clip, width=0.49\textwidth]{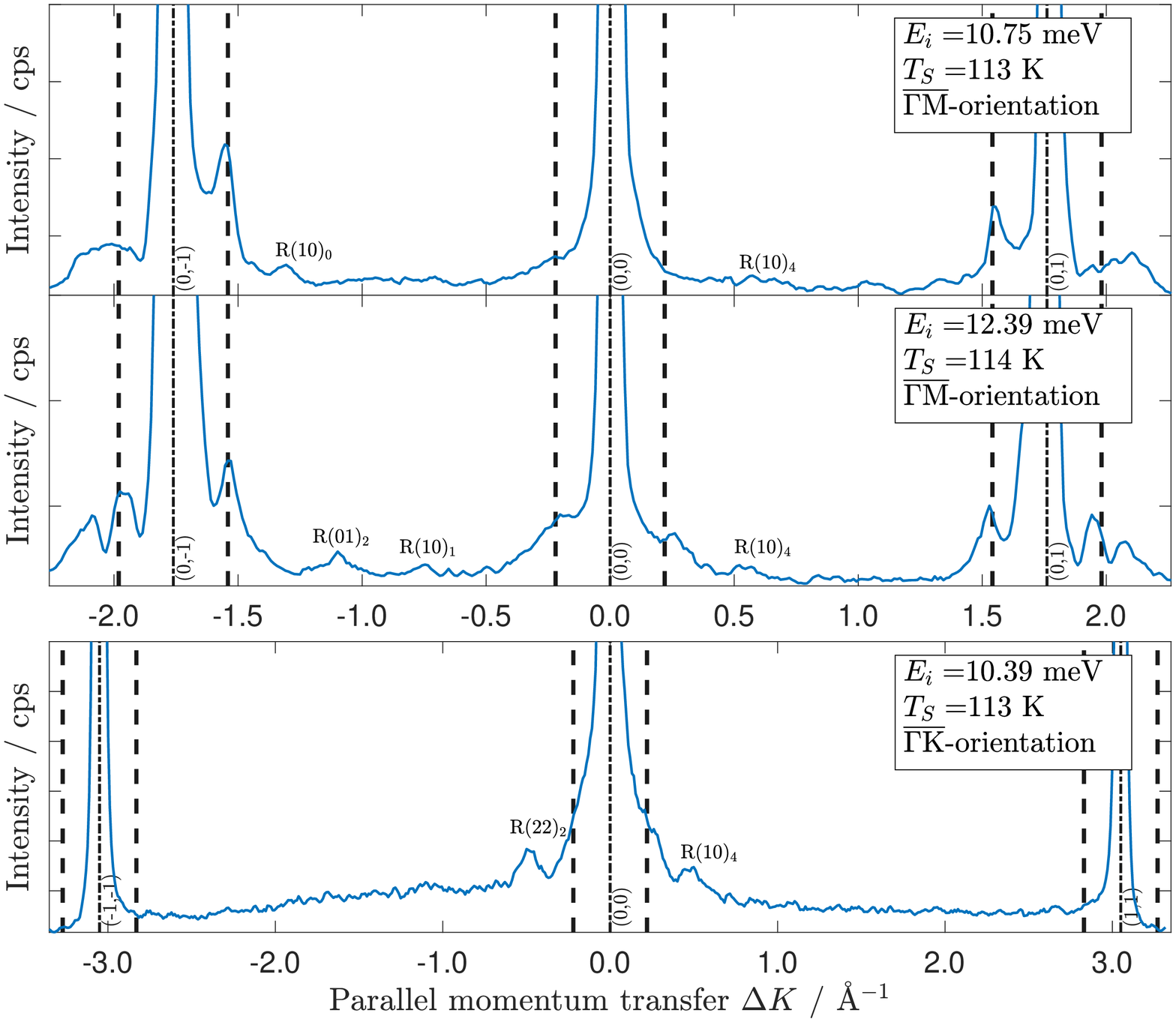}
    \caption{HAS diffraction scans for \Sa (111) measured at low temperature along the two symmetry directions \GM\ and \GK\ . The smaller features between the specular and the two diffraction peaks are due to bound-state resonances and kinematical focusing\cite{Ruckhofer2019}.}
    \label{fig:diffractionSI}
\end{figure}

Despite the diffraction scans along \GM\ which are already shown in the main part of the article, \autoref{fig:diffractionSI} shows also a diffraction scans along the \GK\ azimuth. Note that along \GK\ there is no evidence for additional peaks close to the diffraction peaks, despite two small features close to the specular reflection which can be assigned to resonances\cite{Ruckhofer2019}. This is indicate of the hexagonal shape of the quantum well states giving rise to a multitude of connecting vectors with similar momentum transfer between the flat sides of the hexagon and thus along the \GM\ azimuth.\\

\section{TOF Measurements and Surface Phonon Dispersion}
The phonon energies were determined by performing time-of-flight (TOF) measurements over a wide range of incident angles between the first-order diffraction peaks and at various beam energies. The phonon dispersion was then obtained by calculating the parallel momentum transfer $|\Delta \mathbf{K}|$ for each extracted phonon energy $\Delta E$ from the conservation laws of energy and parallel momentum providing the so-called scan curve for planar scattering:\cite{Tamtoegl2013,Safron1996} 
\begin{equation}
\frac{\Delta E}{E_i} + 1 = \frac{\sin^2 \vartheta_i}{\sin^2 \vartheta_f} \left( 1 + \frac{\Delta \mathbf{K}}{\mathbf{K}_i} \right)^2 
\label{eq:scancurveSI}
\end{equation} 
where $E_i$ is the energy of the incident beam, $\mathbf{K}_i$ is the parallel component of the incident wavevector, and $\vartheta_i$ and $\vartheta_f$ are the incident and final angle, respectively.\\
\begin{figure}[htb]
	\centering
	\includegraphics[width=0.48\textwidth]{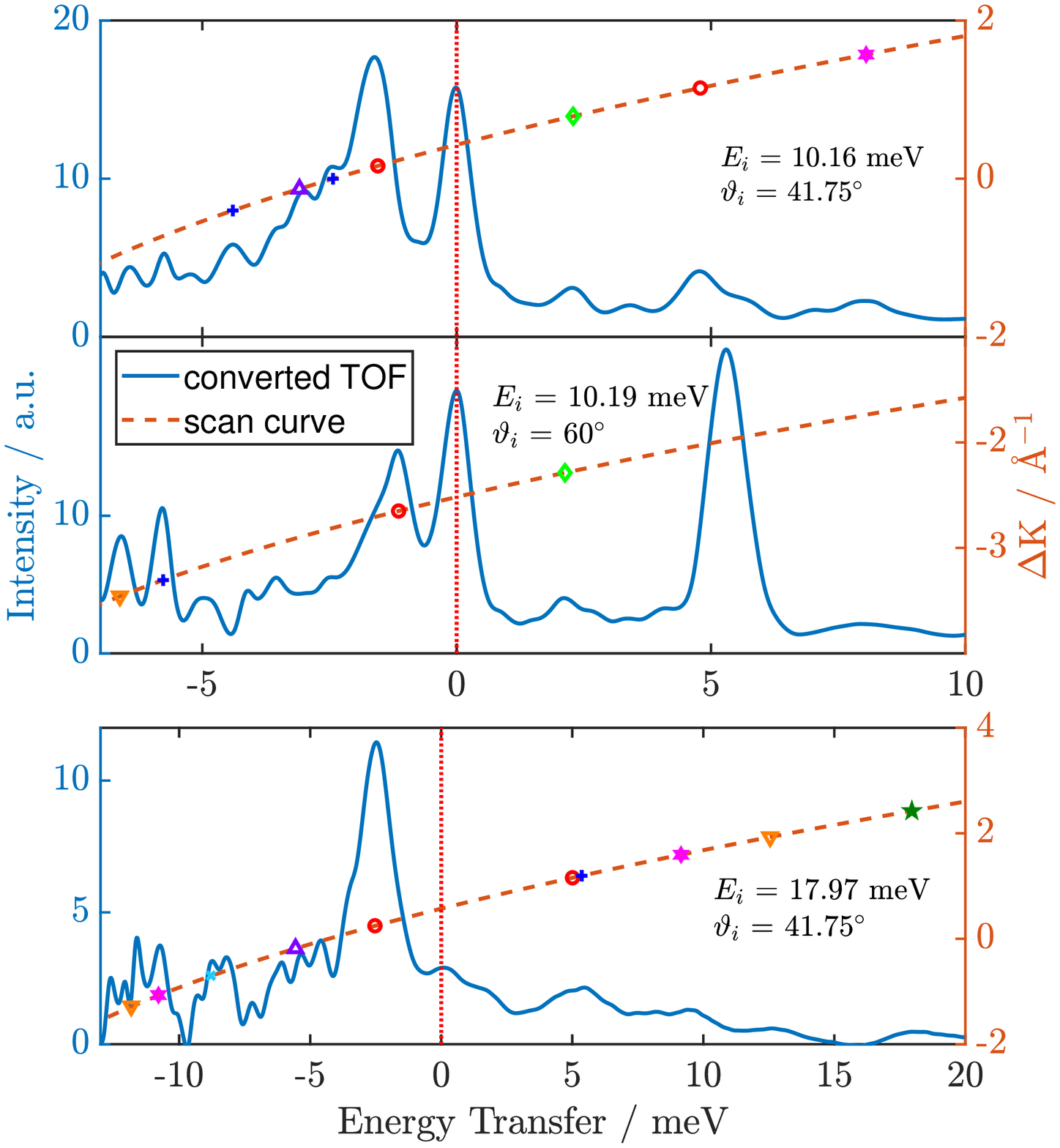}
	\caption{Comparison of different TOF spectra in the $\overline{\Gamma \text{M}}$ direction with various incident Energies $E_i$ and angles $\vartheta_i$. For the two upper graphs the nozzle temperature was set to \SI{50}{\K} and for the bottom one to \SI{80}{\K} while the sample was held at room temperature. The vertical red dotted line corresponds to elastically scattered atoms. The dashed line describes the scan curve which connects energy transfer with momentum transfer ($y$-axis on the right-hand side). The symbols denote peaks in the TOF spectrum which have been assigned to phonon events (same symbols as the main part of the article).}
	\label{fig:Tofcomparison}
\end{figure} 
\autoref{fig:Tofcomparison} shows some further examples of measured TOF spectra along the \GM\ azimuth and the sample at room temperature. Each TOF spectrum consists of various peaks which are located on the creation (negative $x$-axis, $\Delta E<0$) as well as the annihilation (positive $x$-axis, $\Delta E>0$) side. The peak at zero energy transfer corresponds to elastically scattered helium atoms due to the diffuse elastic peak which is caused by scattering from surface imperfections such as steps\cite{Farias1998,Tamtogl2018a}. The scan curve associates each phonon event with a specific momentum transfer $\Delta K$ based on (\autoref{eq:scancurveSI}), since the incoming helium atom looses or gains momentum via inelastic scattering from a phonon.\\
\autoref{fig:Tofcomparison} shows an additional set of TOF-spectra along the $\overline{\Gamma \text{M}}$ direction. Most measurements covering the low energy region with the acoustic phonon modes where performed with a $\approx 10~\mbox{meV}$ beam (top and centre panel in \autoref{fig:Tofcomparison}), while for the optical phonon modes higher incident beam energies (bottom panel in \autoref{fig:Tofcomparison}) where used, in order to cover the optical energy region on the creation side. Note that the particularly strong peak at around $\SI{5.5}{\eV}$ in the centre panel is not assigned to a phonon event. This seemingly inelastic feature originates from elastic scattering and is caused by the outer tails of the velocity distribution in the incident beam. These so-called deceptons or spurions occur within the vicinity of the diffraction peaks and give rise to a large intensity in the inelastic spectra due to their elastic nature.\cite{allison1981}.\\
\begin{figure}[htb]
\centering
\includegraphics[width=0.4\textwidth]{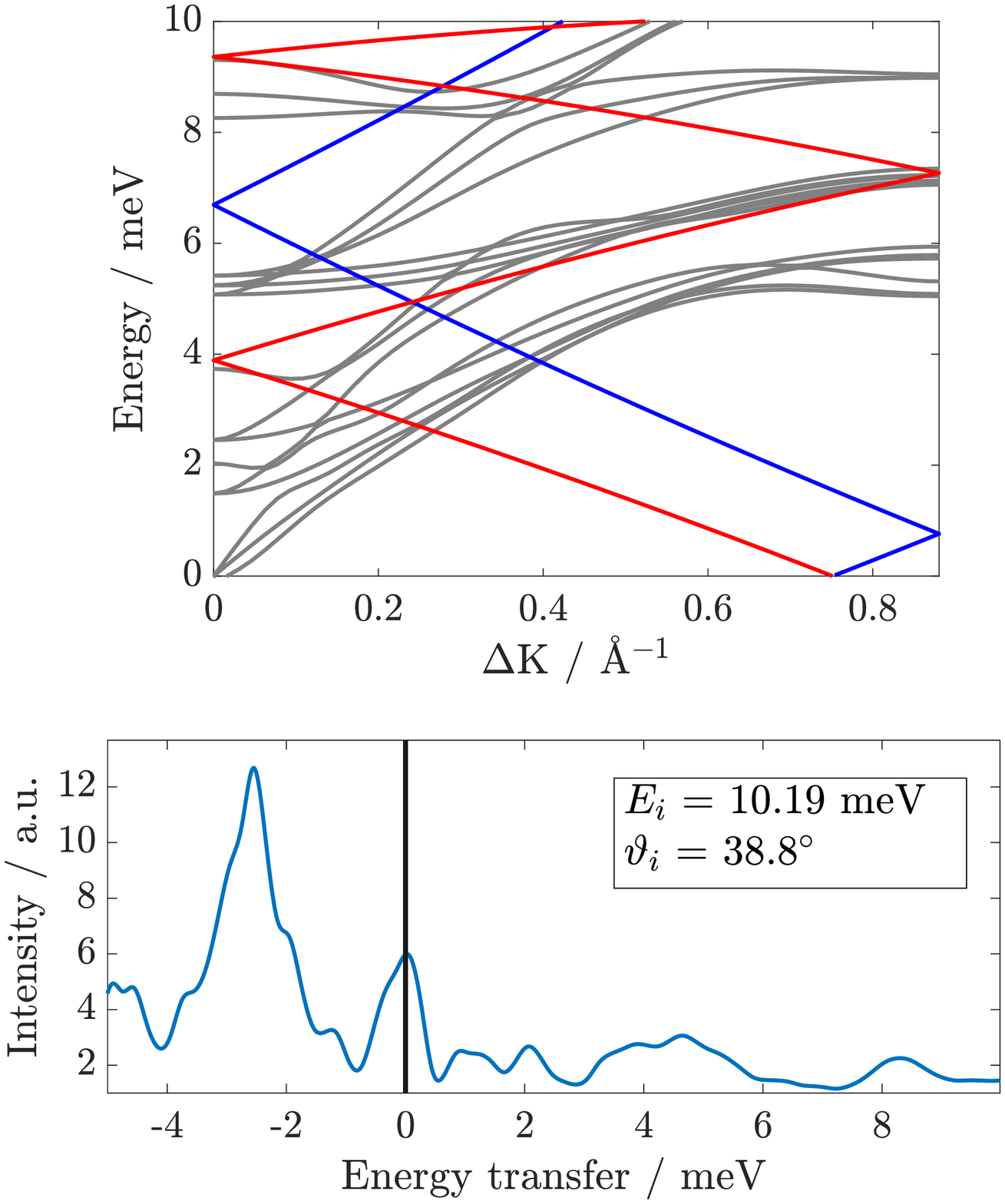}
\caption{The bottom panels shows the TOF spectrum with the $x$-axis converted to energy transfer. The top panel shows the corresponding scan curve (\autoref{eq:scancurveSI}), backfolded into the first Brillouin zone and superimposed onto the results of the DFPT calculation (grey line).}
\label{fig:TOFspectrumSI2}
\end{figure}
As mentioned in the main part of the article, there is no evidence for a strong Kohn anomaly (KA) in our measurements as reported by Zhu \emph{et al}\cite{Zhu2011,Zhu2012}, who observed a strong V-shaped indentation at $\Delta K \approx \SI{0.25}{\AA^{-1}}$ for an optical phonon branch originating at $7.4$ meV at the $\overline{\Gamma}$-point. The scan curve on the phonon annihilation side (blue curve) in \autoref{fig:TOFspectrumSI2}(a) covers exactly the $6-8~\mbox{meV}$ close to the $\overline{\Gamma}$-point. Neither the DFPT calculations (grey lines in \autoref{fig:TOFspectrumSI2}(a)) support such a KA nor does the experimental data which in turn even shows a minimum of the scattered inelastic intensity in the $6-8~\mbox{meV}$ region of the $\Delta E > 0$ side (\autoref{fig:TOFspectrumSI2}(b)).

\end{document}